\documentclass[%
 reprint,
 superscriptaddress,
 amsmath,amssymb,
 aip,
 jap,
]{revtex4-1}

\usepackage{graphicx}
\usepackage{dcolumn}
\usepackage{bm}
\usepackage{hyperref}
\usepackage[version=4]{mhchem}

\begin{document}





\title{Quick-start guide for first-principles modelling of semiconductor interfaces}

\author{Ji-Sang Park}
\affiliation{Department of Materials, Imperial College London, Exhibition Road, London SW7 2AZ, UK}

\author{Young-Kwang Jung}
\affiliation{Department of Materials Science and Engineering, Yonsei University, Seoul 03722, Korea}

\author{Keith T. Butler}
\affiliation{ISIS Facility, Rutherford Appleton Laboratory, Harwell Oxford, Didcot, Oxfordshire OX11 0QX, UK}

\author{Aron Walsh}
\email{a.walsh@imperial.ac.uk}
\affiliation{Department of Materials, Imperial College London, Exhibition Road, London SW7 2AZ, UK}
\affiliation{Department of Materials Science and Engineering, Yonsei University, Seoul 03722, Korea}

\date{\today}

\begin{abstract}
Interfaces between dissimilar materials control the transport of energy in a range of technologies including solar cells (electron transport), batteries (ion transport), and thermoelectrics (heat transport). Advances in computer power and algorithms means that first-principles models of interfacial processes in realistic systems are now possible using accurate approaches such as density functional theory. In this `quick-start guide', we discuss the best practice in how to construct atomic models between two materials and analysis techniques appropriate to probe changes in local bonding and electronic band offsets. A number of examples are given related to perovskite solar cells.
\end{abstract}

\maketitle



\textit{`The interface is the device'} stated Herbert Kroemer in his Nobel lecture.\cite{H}
The behaviour of materials interfaces encompasses a range of properties and processes, which can include:
\begin{itemize}
\item electronic band alignment, influencing the transport and confinement of electrical charge;
\item mechanical strain, altering the local electronic structure;
\item chemical bonding environment, resulting in new electronic states active for trapping electrons and/or ions;
\item accumulation of intrinsic point defects and impurities, with enhanced mobility of defect centres.
\end{itemize}
The development of microscopic models for materials interfaces has a rich history. For example, in 1935 Gurney studied the effect of alkaline earth metals on the workfunction of tungsten using quantum mechanics\cite{Gurney1935a}. 
Knowledge of valence and conduction band offsets contributed to the development of rectifiers and transistors.\cite{Mott1939a} 
{\color{red} In the late 1970s, the first descriptions of interfaces based on self-consistent density functional theory (DFT) were reported\cite{Pickett1977,Louie1976}. 
Procedures for the calculation of electronic band offsets were suggested later,\cite{Resta1989,VandeWalle1989} and were widely applied to compound semiconductors in the 1990s.\cite{Wei1998,Peressi1998}
}

{\color{red} Beyond semiconductor-semiconductor interfaces, a wide range of solid interfaces have been studied including those involving metals, semiconductors and insulators, partly motivated by their importance in transistor technology.\cite{Louie1976,Pasquarello1998,robertson2005interfaces,Stengel2007,ruiz2012density,li2013transparent}}
The latest generation of materials employed in energy storage and conversion -- including perovskites, MXenes, spinels, olivines -- are  multicomponent and adopt crystal structures more complex than face-centred-cubic (fcc) semiconductors.\cite{Catlow2010a} 
{\color{red} Investigation of realistic interfaces in these technologies (e.g. thermoelectrics, batteries, power electronics) pose scientific challenges.\cite{Tian2012,Hu2013,Abavare2013,walsh2018taking}}
Here, we discuss how to approach modelling and analysing interfacial processes using first-principles computational approaches.    

\begin{figure}[t]
\includegraphics[scale=0.42]{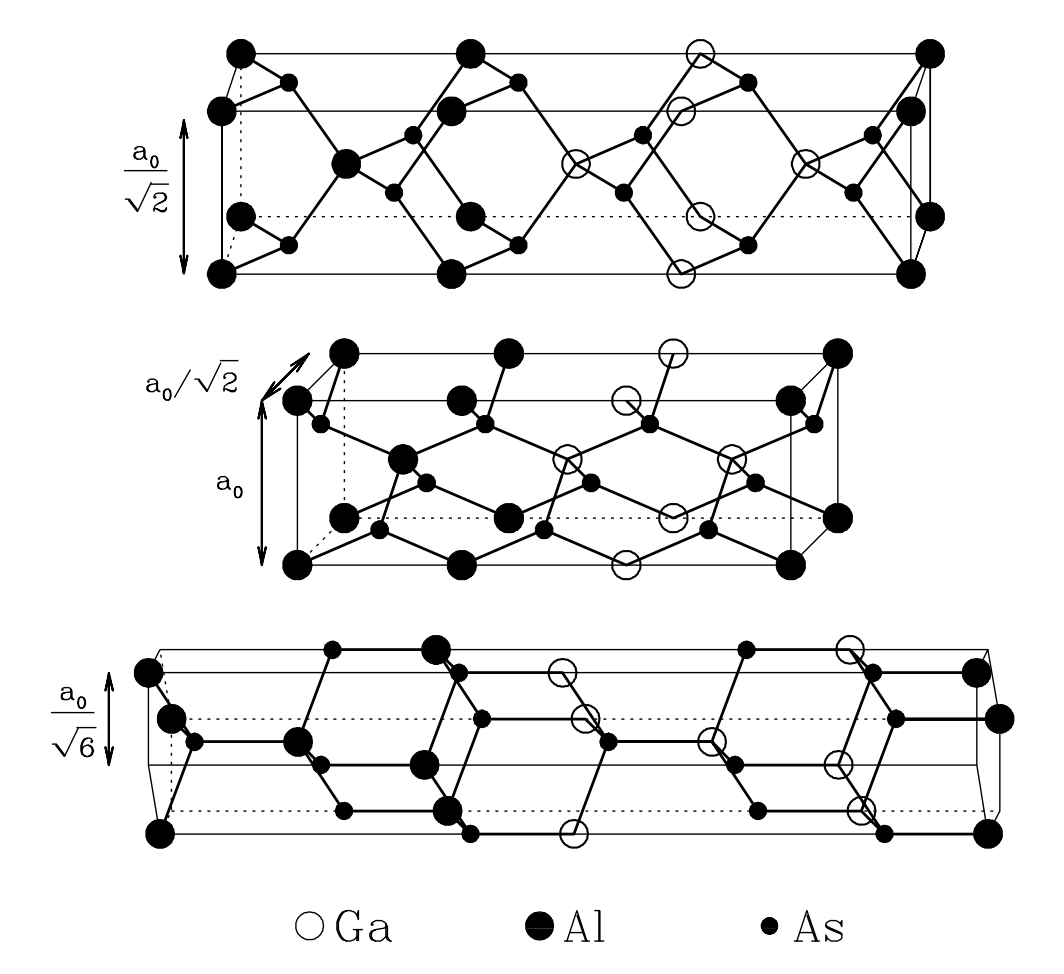}
\caption{\label{f-bald} Typical supercell expansions used to build interfaces between two zincblende (fcc) structured semiconductors for the example of
\ce{GaAs | AlAs} (001), (110) and (111) abrupt junctions (from the top to the bottom).
Reproduced from Ref. \onlinecite{Peressi1998}.
} 
\end{figure}

\section{Isostructural interface models}

A special case for building interface models is when both materials share the same structure type.
An atomic model can be readily constructed along different crystal orientations
by a simple transformation of the unit cell.
For example, a cubic unit cell is reoriented along $<111>$ using a rotation matrix of:
\begin{equation}
 \begin{bmatrix}
   -1  & -1/2   & 1 \\
   1  & -1/2 & 1        \\
   0  & 1 & 1        \\
\end{bmatrix} 
\end{equation}
An example interface for fcc semiconductors is illustrated in Figure \ref{f-bald}. 
A particular orientation may be favoured thermodynamically, following the most stable crystal surfaces, or may be determined by the fabrication procedure of a device. 
Relating to thin-film solar cells, both \ce{CH3NH3PbI3} and \ce{CH3NH3PbBr3} adopt a perovskite-type structure with a low lattice constant mismatch ($\frac{\Delta a}{a} < 3 \%$),
which can be used to construct small ($<$ 100 atom) supercell models of the (001) interface that are practical for DFT calculations.\cite{Butler2015}

\begin{figure}
\includegraphics[scale=0.475]{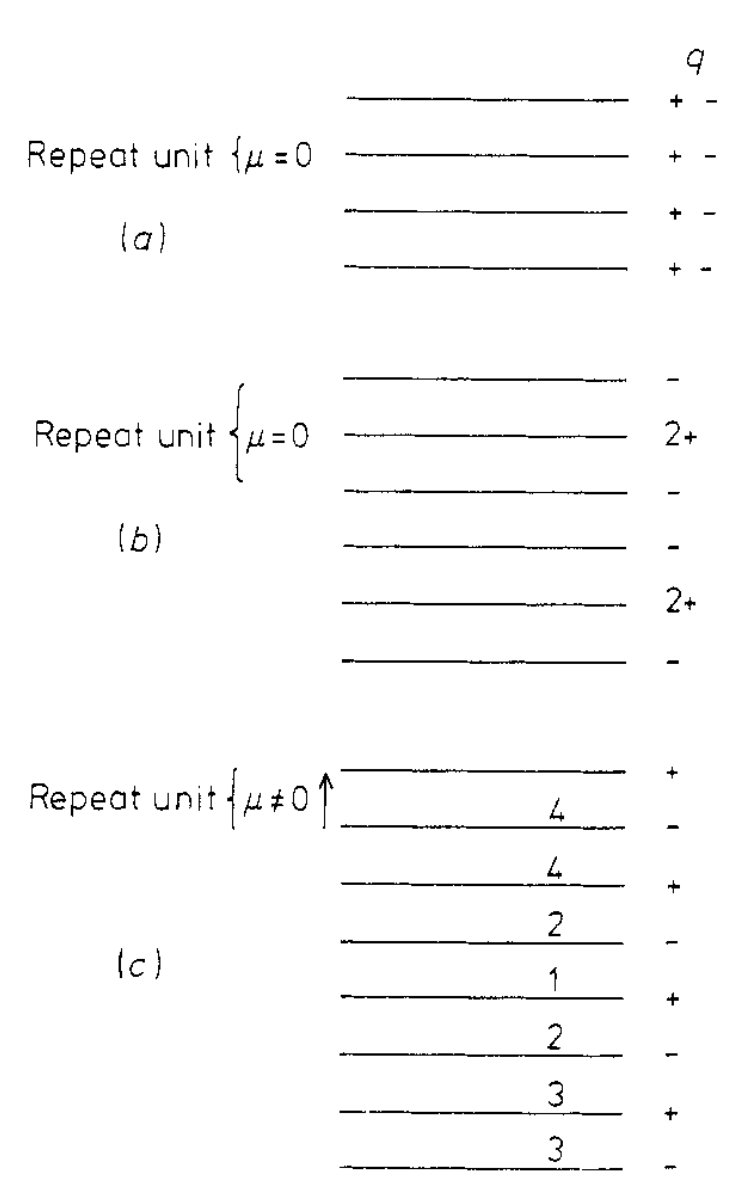}
\caption{\label{f-tasker} Classification of surface terminations by Tasker.
Distribution of charges $q$ on planes for three stacking sequences parallel to surface. (a) Type 1 (\textit{non-polar}) with equal anions and cations on each plane; (b) Type 2 (\textit{quadrupolar}) with charged planes but no net dipole moment ($\mu$); (c) Type 3 (\textit{dipolar}) charged planes and dipole  normal to the surface.
Reproduced from Ref. \onlinecite{Tasker1979}.
} 
\end{figure}

For ionic materials, care is needed as a particular crystal orientation may carry an electric dipole moment.\cite{Evjen1932}
The classification of polar surfaces by Tasker\cite{Tasker1979} is summarised in Figure \ref{f-tasker}.
For example, the (100) surface of rocksalt is non-polar (Type 1), the (110) surface is quadrupolar (Type 2), and the (111) surface is polar (Type 3).

There may be motivation to study a specific interface orientation (e.g. from microscopy), but in general a non-polar orientation should be chosen to avoid complications in practical calculations. 
Type 3 surfaces generate a macroscopic electric field that is proportional to the size of the supercell expansion. 
In real materials, such an electric field is removed by structural or chemical reconstructions, but in calculations it can cause poor convergence in properties with respect to slab thickness, and unphysical behaviour such as charge `sloshing' between the two sides of the interface. 

\section{Heterostructural interface models}
For interfaces involving two materials with different crystal structures, challenges arise owing to the choice in the relative orientations of the two crystals. Often, experimental information on the preferred morphology is not available, especially for newer technologies. 

\begin{figure}[t]
\includegraphics[scale=0.4]{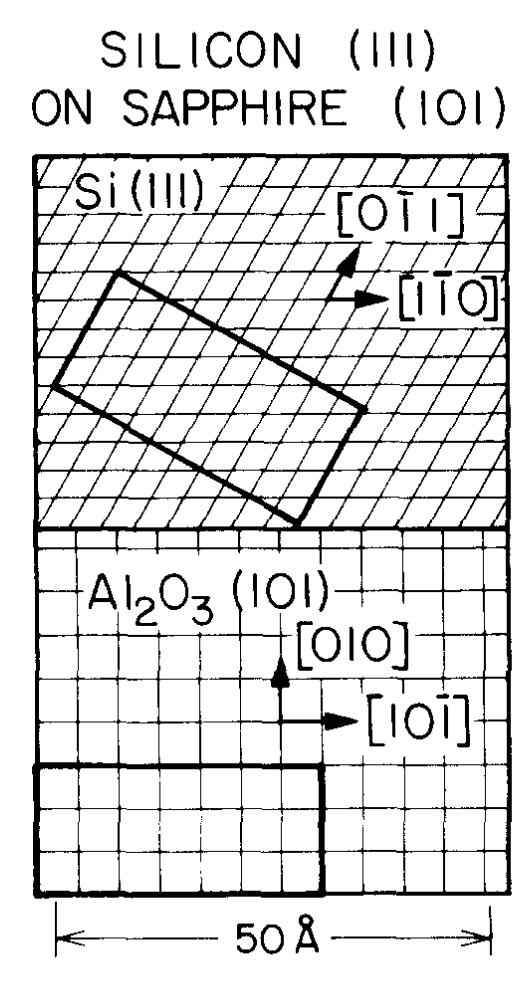}
\caption{\label{f-zur} Lattice translations parallel to the Si(111) and sapphire (101) faces. A cell made of 21 sapphire unit cells has almost the same dimensions as a cell made of 40 Si unit cells. \ce{Al2O3} is rhombohedral with $a$ = 5.129 \AA, and $\alpha$ = $55^\circ 17'$. Silicon is face-centered cubic with $a$ = 5.431 \AA. Reproduced from Ref. \onlinecite{Zur1984}.
} 
\end{figure}

Zur and McGill\cite{Zur1984} proposed an algorithm \textsc{lattice match} to identify suitable periodic reconstructions of an interface between two dissimilar materials. A search is performed of lattice directions and expansions, and they are reduced to their simplest form. 
The example of Si on \ce{Al2O3} is shown in Fig. \ref{f-zur}.
The approach was recently implemented by Butler \textit{et al} to predict lattice-matched electrical contacts suitable for perovskite solar cells, which identified a favourable match between \ce{CH3NH3PbI3} and \ce{Cu2O}.\cite{Butler2016a}
{\color{red} Other procedures for lattice matching have also been reported.\cite{lazic2015cellmatch,jelver2017determination}}

While procedures such as \textsc{lattice match} can identify low-strain orientations, the underlying chemical bonding (atomic sites and bonding) is neglected.
Coincident Site Lattice Theory addresses the atomistic detail.\cite{Brandon2010}
At certain misorientation angles, there may be more coincident lattice sites (i.e. a common sublattice) in addition to the origin, and such configurations can be stabilized over others with a similar misorientation angle. 
One might expect that an interface model with a high planar density of coincidence sites to be most stable; however, this strongly depends on how the atoms are bonded at the interface.
For example, in an ionic material cations prefer to be in contact with anions.\cite{Sutton1987}

In a recent study of interface formation between perovskite-structured \ce{CsPbBr3} and rocksalt-structured \ce{PbS} -- a system with potential applications in light emitting diodes --
a number of high-coordination configurations were identified and their stability assessed.\cite{Jung2017a} 
The interface configuration where Pb of \ce{CsPbBr3} is located on S of \ce{PbS}, and Br of \ce{CsPbBr3} is located on Pb of \ce{PbS}, was found to be most feasible due to favourable electrostatic interactions that lower the interface energy.

\section{Interface analysis}

Once a reliable atomic model of an interface between materials $X$ and $Y$ has been constructed, and calculations have been successfully performed, properties can then be analysed. 
While some application areas require specific analysis -- such as the potential energy barriers for ion diffusion across an interface in solid-state batteries\cite{ORourke2018} --
the points discussed here are intended to be general. 

\subsection{Interface energy}
Concerning thermodynamic stability, similar to the calculation of a defect formation energy,\cite{park2018point} the interface energy $E_{f}(X/Y)$ can be calculated from the general expression:
\begin{equation}
    E_{f}(X/Y) + E_{f}(Y/X) = ( E_{tot}(X/Y) - \Sigma_{i} n_i \mu_i ) / A
\end{equation}
where $E_{tot}(X/Y)$ is the total energy of interface model, $A$ is the area of the interface in the supercell model, and $\mu_i$ is the chemical potential of atomic species $i$.
{\color{red} Since materials $X$ and/or $Y$ are strained biaxially in the interface calculations, the reference chemical potential is usually obtained from the strained bulk materials, and it is also important to consider relaxation in the direction normal to the strain plane (i.e. the Poisson effect).}
{\color{red} The chemical potential can also be used to account for different growth conditions, including contributions from off-stoichiometry and impurities.}

Owing to periodic boundary conditions, two interfaces are usually contained within a supercell. 
Both $X/Y$ and $Y/X$ interfaces must be identical to obtain a well-defined interface energy for a particular interface; this corresponds to a supercell model with inversion symmetry. 
{\color{red} If the two interfaces are not identical, one can employ a vacuum slab geometry to obtain the interface energy for each interface in turn,
while properly accounting for the associated surface energy.
}

If the models of $X$ and $Y$ are stoichiometric, 
then the chemical potential can be removed and the interface energy expression is simplified to:
\begin{equation}
    E_{f}(X/Y)= \frac{E_{tot}(X/Y) - n_{X} E_{tot}(X) - n_{Y} E_{tot}(Y)}{2A} 
\end{equation}
where $n_{X}$ is number of stoichiometric units of $X$ and $E_{tot}(X)$ is the total energy of bulk $X$.
Values of interface energy typically range from 0--5 Jm$^{-2}$ (0.3 eV/\AA$^2$).\cite{Tran2016,Hemmingson2017,Scheiber2016}

Otherwise, the interface stability can be inferred from the work associated with placing the surfaces of $X$ and $Y$ in contact:
\begin{equation}
    W_{sep}(X/Y)= \frac{E_{tot}(X/Y) - E_{slab}(X) - E_{slab}(Y)}{2A}
\end{equation}
where $W_{sep}(X/Y)$ is the work of separation, $E_{slab}(X)$ and $E_{slab}(X)$ are the total energy of isolated $X$  and $Y$ surface slab models. 
Therefore, negative $W_{sep}(X/Y)$ represents a preference for $X$ and $Y$ to form an interface. 

$W_{sep}$ is readily obtained from calculations, but in experiment cleavage of an interface also results in elastic, reconstructive, and diffusive processes that act to lower the energy of the cleaved surfaces. 
The measured quantity is called the work of adhesion $W_{ad}$ and differs from $W_{sep}$ by varying degrees;
comparison between the two quantities should be made with caution. 
$W_{sep}$ is relevant for mechanical properties of interfaces; however, for dynamical processes such as surface wetting differences between $W_{sep}$ and $W_{ad}$ can be important.\cite{Finnis1996}

\subsection{Structure change}
Interface simulations can be used to probe  changes in bond lengths and angles across an interface,
simply by inspection of the atomic coordinates. 
Due to the abrupt change in coordination environments across an interface, the differences before and after geometry optimisation can be large. 

In addition to the equilibrium structure, the impact of tensile/compressive strain can be also investigated. 
If material $Y$ is grown pseudomorphically on material $X$, then $Y$ experiences biaxial  strain, depending on the difference in lattice constant. 
The physical properties (e.g. band gap, point defect formation, and ion diffusivity) will be affected accordingly. 
Crystal strain could be beneficial, e.g. as employed in Si electronics where the channel Si is intentionally strained to increase the carrier mobility.\cite{Sun2007a}
Strain at an interface can be reduced by elemental intermixing,\cite{Hybertsen1990}
e.g. by having a $X_{1-y}Y_{y}$ compositional gradient rather than an abrupt junction;
however, this can be challenging to model using atomistic simulation techniques due to the large model systems required.

\begin{figure}[b]
\includegraphics[scale=0.25]{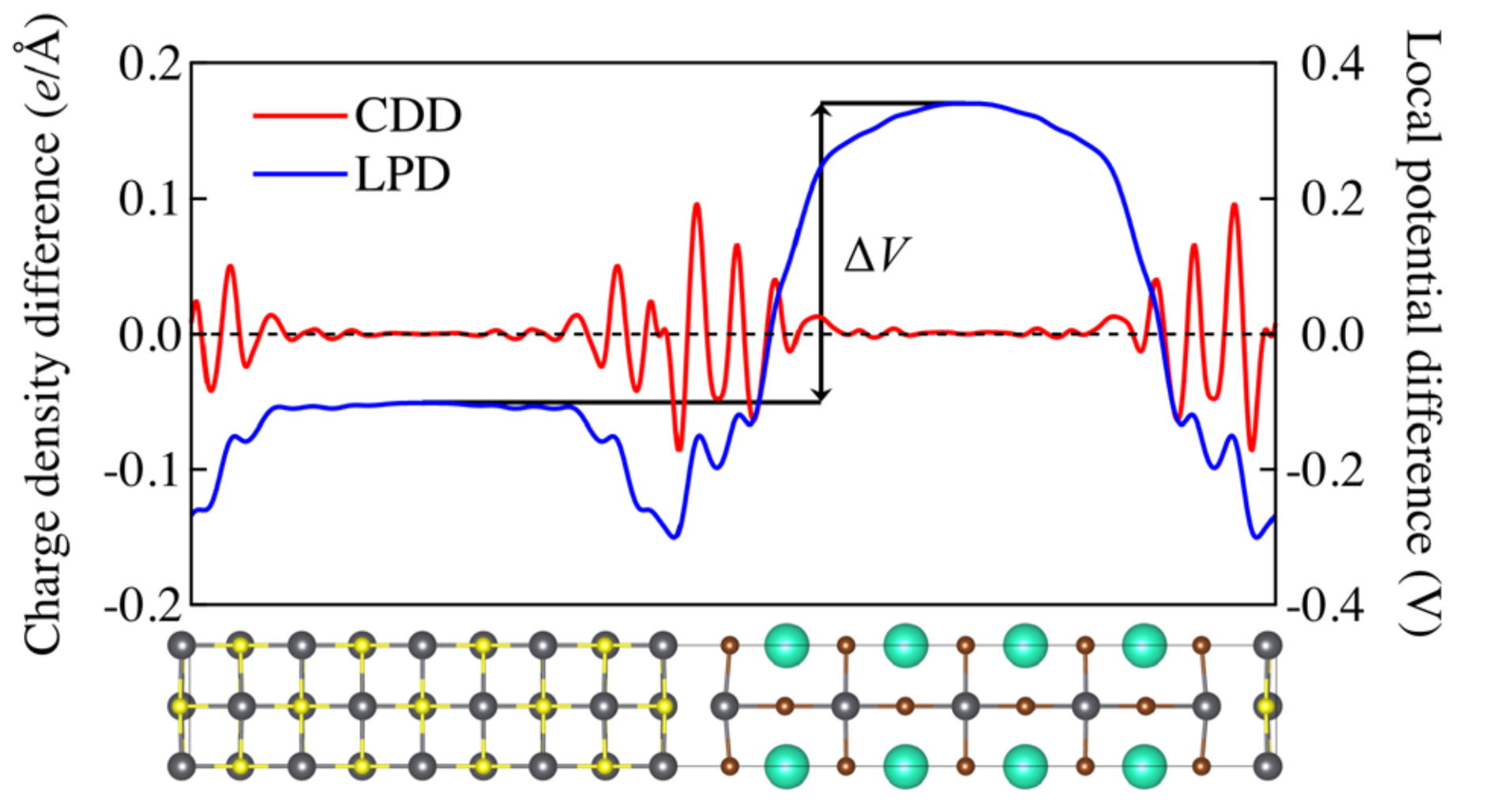}
\caption{\label{f-yk} 
Calculated properties of a \ce{PbS | CsPbBr3} (001) heterojunction. 
Planar-averaged charge density difference (red line, labeled as CDD) and local potential difference (black line, labeled as LPD). Potential difference ($\Delta$V) is measured between two plateaus at the center of each material. 
Reproduced from Ref. \onlinecite{Jung2017a}.
} 
\end{figure}

\subsection{Electrostatic potential}
Beyond the local structure, the behaviour of mobile charge (ions and electrons) is influenced by the changes in electrostatic potential towards an interface.
An example for the \ce{PbS | CsPbBr3} interface is shown in Fig. \ref{f-yk}.
The 3D electrostatic potential fluctuates rapidly because of the atomic potentials. 
To obtain plateaus useful for alignments, the potential can be averaged over the supercell (length $\Delta$) along the interface normal direction ($z$) as follows:\cite{Resta1989}
\begin{equation}
    \overline{V}(z) = \frac{\int \int \int_{z-\Delta/2}^{z+\Delta/2} V(x,y,z') dz' dy dx}{\int \int \int_{z-\Delta/2}^{z+\Delta/2} dz dy dx}
\end{equation}
Plateaus are obtained in bulk-like regions if the dipoles at each interface cancel out.\cite{Peressi1998} 
Tools such as \textsc{Macrodensity}\cite{Butler2014} allow for a range of potential averaging methods to be tested.

\begin{figure}
\includegraphics[scale=0.3]{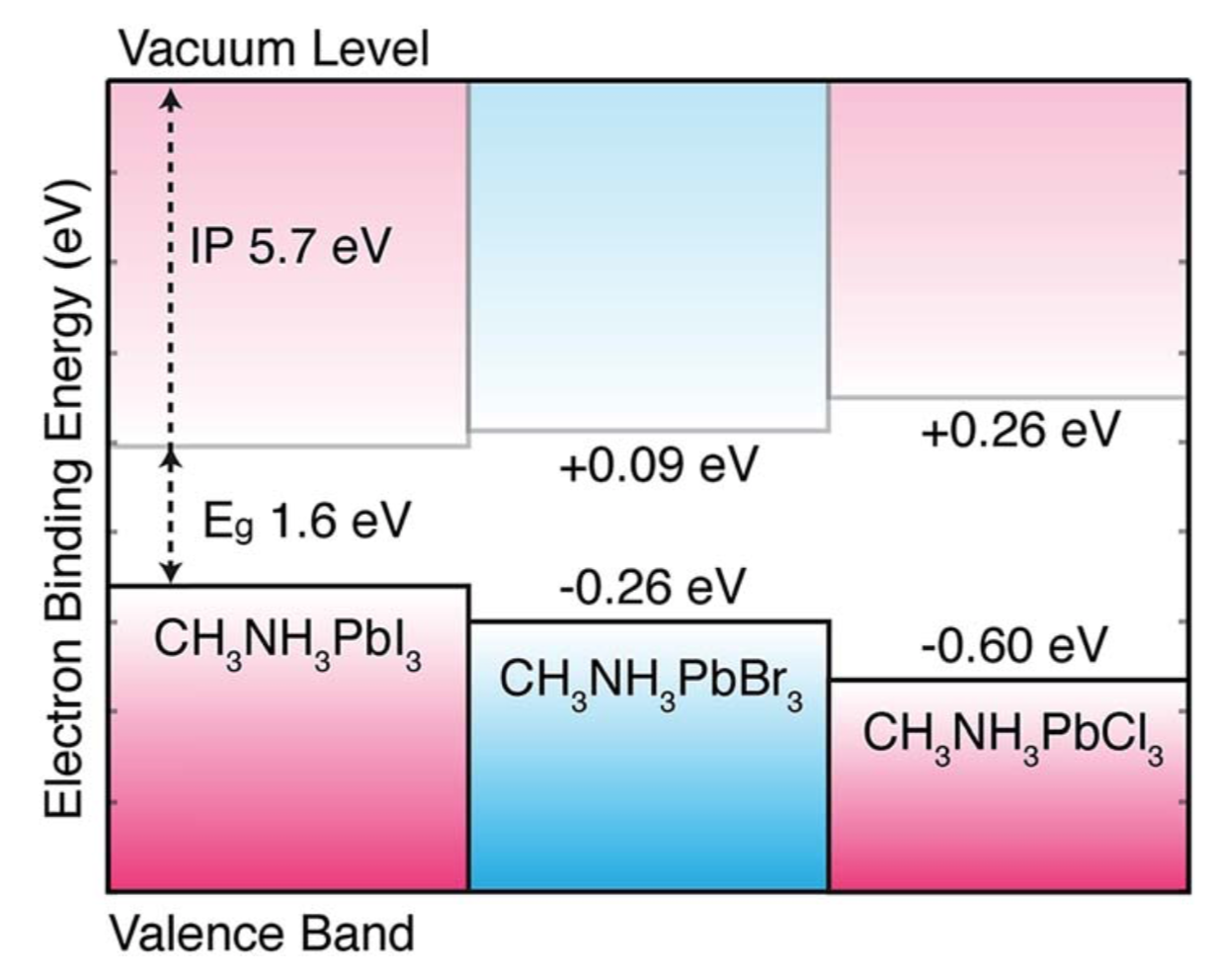}
\caption{\label{f-halides} 
Calculated electronic band alignment of three halide perovskite semiconductors using a core-level alignment procedure using Pb 1s (see Ref. \onlinecite{Li2009a} for further detail).
The energies are given with respect to the band edge positions of \ce{CH3NH3PbI3} including corrections from quasi-particle self-consistent \textit{GW} theory.
Reproduced from Ref. \onlinecite{Butler2015}.
} 
\end{figure}

The electronic band edge positions of the bulk materials that form interface can be obtained using an appropriate alignment procedure; a selection of different approaches can be found in Refs.
\onlinecite{Anderson1960,Tersoff1984a,Jaros1988,Peressi1998,VandeWalle1989,Wei1998,Li2009a,Monch2011,Kumagai2017}.
There are similarities to calculating the workfunction (or ionisation potential) of an isolated surface.  
An example is shown in Figure \ref{f-halides} for the \ce{CH3NH3PbI3 (X = Cl, Br, I)} series.\cite{Butler2015}
The valence and conduction band offsets were calculated using an alignment procedure that employs deep-lying core states, with results equivalent to the raw electrostatic (Hartree) potential. 
From this analysis, the band offset class can be assigned to distinguish between the cases of charge confinement (`straddling', type I), charge separation (`staggered', type II), or a conductive junction (`broken gap', type III). 
{\color{red} We note that the natural band offset procedure described above is not suitable for observing local band gap narrowing or widening at an interface, which can be analysed by the local density of states (DOS).\cite{Crovetto2017}}

\subsection{Electronic structure}
In standard calculations, the electronic DOS is  summed over the entire simulation cell; however, it can be projected in real space along the interface axis to probe changes in chemical bonding towards the junction. 

Analysis of the charge density difference ($\Delta \rho = \rho_{(X|Y)} - \rho_X - \rho_Y$) can be used to show how electrons are redistributed upon interface formation.
If alternating charge accumulation and depletion regions are shown along the contact area, this implies chemical bonding that enhances interface stability. 
On the other hand, if a charge accumulation (depletion) region is shown alone, this could imply cation-cation (anion-anion) repulsion across a polar interface. 

The magnitude of the electronic band gap can be increased or decreased in interface calculations because of strain, as mentioned above. Metal-induced gap states (MIGS) can also be introduced in the semiconductor side.\cite{Tersoff1984a} The MIGS result in charge redistribution and modified offsets at the interface.\cite{Monch1996a} 
DFT simulations with spatially resolved DOS can be used to characterise MIGS at metal--semiconductor interfaces.\cite{Profeta2001}

At metal--semiconductor interfaces, an additional consideration is the image-potential.\cite{Inkson1973,Arita2001} 
In 1983, Stoneham proposed that the image potential could be the dominant contribution to surface adhesion and metal support effects in catalysts, where metal ceramic interfaces are present.\cite{Stoneham1983} 
Later \textit{ab initio} calculations of metal-ceramic interfaces confirmed that the image-potential is a leading term in determining interface stability in such systems.\cite{Finnis1992}

{\color{red}
Interface properties can also be investigated using the non-equilibrium Green's function (NEGF) formalism, which is often used for electron transport. 
A non-periodic interface is achieved by introducing semi-infinite electron reservoirs on each side of the simulation cell. 
Another advantage of this approach is that current-voltage characteristics can be compared directly with  experiment.\cite{Crovetto2017}}

\section{Convergence of calculations}

There are standard convergence criteria for electronic structure calculations of bulk solids, which can include changes in total energy, ionic forces, and structure parameters with respect to the \textit{k}-point sampling and the quality of basis set. 
The main two additional factors for interface calculations are slab orientation and thickness,
which need to be converged with respect to the property of interest.
{\color{red}
The description of electron transfer and space charging effects may be particularly sensitive to the supercell size, which is typically much smaller than the physical screening length for most materials.
}

There is also the choice of electron exchange and correlation functional in DFT. While one functional may provide a reliable description of bulk properties of one material, at the interface the chemical bonding is perturbed and the electronic wavefunction can be more localised. 
An accurate description of interfacial bonding may require hybrid functionals or beyond, depending on the chemistry of the system being studied.\cite{Alkauskas2008} 

When calculating separate representations of the bulk, surface and interface systems, it is advisable to check that all simulation cells are in the same crystallographic orientation. 
For reasons of numerical accuracy, ensuring that all cells have the same internal parameters (e.g. grid density and \textit{k}-point sampling) leads to smoother convergence of the terms in Equations 2 -- 4. 

For example, if simulating an interface along the $<111>$ direction, the transformation in Equation 1 should be applied to the bulk as well as to the interface model. 
Additionally, the vacuum spacing in the surface model should be chosen to be an integer supercell expansion of the bulk to enable to clean comparison between the surface and interface models. 

\section{Conclusion}

Procedures for simulating bulk crystalline materials are well-established. 
The combination of two or more materials to form an interface raises additional challenges for materials modelling, 
but can yield valuable insights that are difficult to probe experimentally. 
We have shown how consideration of the surface terminations and lattice-matching conditions are useful tools to generate interface models. 
Furthermore, we discussed and provided introductory references on how to approach analysis of the physical properties including thermodynamic stability, bonding, and electronic structure. 
There are many application areas for these techniques across emerging energy technologies. 

\acknowledgments
The research was funded by the Royal Society and the  EU Horizon2020 Framework (STARCELL, grant no. 720907).
Additional support was received from the Faraday Institution (grant no. FIRG003).

\bibliography{library}

\end{document}